\documentclass[twocolumn,aps,prl]{revtex4-1}
\usepackage{amsmath,amssymb}
\usepackage{graphicx}
\usepackage{epstopdf}
\usepackage[usenames]{color}
\usepackage{array} 
\usepackage{mathtools}
\usepackage{braket}
\usepackage{subfig} 
\usepackage{caption}
\usepackage{multirow}

\newcommand{\be}{\begin{equation}}
\newcommand{\ee}{\end{equation}}
\newcommand{\bk}{{{\bf{k}}}}

\newcommand{\bea}{\begin{eqnarray}}
\newcommand{\eea}{\end{eqnarray}}
\newcommand{\ra}{\rangle}
\newcommand{\la}{\langle}

\newcommand{\upa}{\uparrow}
\newcommand{\dna}{\downarrow}

\newcommand{\dg}{{\dagger}}
\newcommand{\pdg}{{\phantom\dagger}}

\newcommand{\dk}{d \kern-0.25em \bar~ k}
\newcommand{\dw}{d \kern-0.25em \bar~\omega}

\let\upspn=\uparrow
\let\dwnspn=\downarrow
\let\goto = \rightarrow


\renewcommand{\vec}[1]{\ensuremath{\mathbf{#1}}} 
 
\newcommand{\abs}[1]{\left| #1 \right|} 
\newcommand{\avg}[1]{\left< #1 \right>} 
 
\let\baraccent=\= 
\renewcommand{\=}[1]{\stackrel{#1}{=}} 

\let\upspn=\uparrow
\let\dwnspn=\downarrow
\let\goto = \rightarrow

\begin{document}

\title{Thermal Phase Transitions of Strongly Correlated Bosons with Spin-Orbit Coupling}
\author{Ciar\'{a}n Hickey$^1$ and Arun Paramekanti$^{1,2}$}
\affiliation{$^1$Department of Physics, University of Toronto, Toronto, Ontario, Canada M5S 1A7}
\affiliation{$^2$Canadian Institute for Advanced Research, Toronto, Ontario, M5G 1Z8, Canada}
\begin{abstract}
Experiments on ultracold atoms have started to explore lattice effects and thermal fluctuations for two-component bosons with
spin-orbit coupling (SOC). Motivated by this,
we derive and study a $tJ$ model for lattice bosons
with equal Rashba-Dresselhaus spin-orbit coupling (SOC) and strong Hubbard repulsion in a uniform Zeeman magnetic field. Using the
Gutzwiller  ansatz, we find strongly correlated ground states with stripe superfluid (SF) order.
We formulate a finite temperature generalization of the Gutzwiller method, and show that thermal
fluctuations in the doped Mott insulator drive a two-step melting of the stripe SF, revealing a wide regime of a stripe normal fluid.
\end{abstract}
\maketitle



Spin orbit coupling (SOC) underlies a diverse range of remarkable phases in solid state materials including topological insulators \cite{KaneHasan_RMP2010,QiZhang_RMP2011}, 
quantum anomalous 
Hall insulators \cite{HaldaneQAH_PRL1988,Chang_Science2013}, and Skyrmion crystals \cite{Tokura2010}, 
while its interplay with strong correlations is expected to lead to exotic topological 
Mott insulators \cite{PesinBalents_NPhys2010, KrempaKim_ARCM2014}. Experiments on ultracold atomic gases have started to explore
analogous issues for SOC in 
Bose fluids, using Raman transitions
to induce an equal Rashba-Dresselhaus SOC and a uniform Zeeman magnetic field 
\cite{Lin2011,Pan2012,Dalibard2013,LJLBlanc_NJP2013,Engels_PRA2013,BeelerQSH_Nature2013,PanFiniteTExpt_NPhys2014, Hammer2014,JZhang_NPhys2014}.
Striking observations include the spin Hall effect \cite{BeelerQSH_Nature2013} and tunable production of Feshbach molecules  
\cite{JZhang_NPhys2014}.
On the theoretical front, Bose
superfluids with equal Rashba-Dressehaus coupling have been shown to exhibit stripe orders, and spin and density coupled collective modes
\cite{Galitski2008,Zhai2010,Ho2011,Radic_PRA2011,LiPitaevskii_PRL2012,MartoneLiPitaevskii_PRA2012,Zhai2012,LiPitaevskii_PRL2013,Sheehy2013}.
Pure Rashba SOC, with a circular minimum in the single particle dispersion, may 
lead to unusual fluctuation effects \cite{Zhai_PRB2011,Baym_PRL2012,Barnett_PRA2012,Baym_PRA2012,Juzeliunas2013}, ferromagnetism \cite{Kopietz2013}, 
or topological ground states \cite{Sedrakyan2012,HPu2013}.
Incorporating strong correlations on a lattice induces superfluids or Mott insulators with remarkable spin textures
\cite{Grass2011,Mandal2012,WSCole_PRL2012,RadicGalitski_PRL2012, ZiCai_PRA2012,SZhang2014,CZhang2013,WMLiu2014,Hofstetter2014}
and topological transport properties \cite{Duine2013}.
Very recently, experiments have started to explore thermal phase transitions \cite{PanFiniteTExpt_NPhys2014},  and 
the effects of a periodic lattice potential \cite{Hammer2014}, in a Bose-Einstein condensate (BEC) with equal Rashba-Dresselhaus
SOC.

Motivated by the broad interest in understanding the interplay of SOC and strong correlations, and ongoing experimental efforts in ultracold gases,
we focus here on two important questions. (a) How does the presence of a lattice and strong correlations 
modify the ground states of bosons with equal Rashba-Dresselhaus SOC? (b) How do thermal fluctuations impact Bose superfluids with SOC?
Our key results are the following. (i) At strong correlations, we derive an effective $tJ$ model for lattice bosons with equal Rashba-Dresselhaus SOC and 
a uniform magnetic field.
Using a zero temperature Gutzwiller ansatz, we show that this leads to strongly correlated variants
of stripe and incommensurate SFs previously discussed in the continuum. However, unlike in the continuum, applying a large magnetic field leads to three 
distinct SFs (see Fig.~1(a,b)) depending on the SOC angle: (a) a zero momentum SF analogous to the continuum case, (b) a 
$\pi$-momentum SF, or (c) a $\pi/2$-momentum SF. (ii) At weaker field, strong interactions induce stripe
order; in contrast to the continuum, the stripe order has significant higher harmonic content resulting in 
extra peaks in the momentum distribution as seen from Fig.~1(c,d). (iii) Previous work has considered
thermal fluctuations of weakly interacting continuum
bosons with SOC \cite{Zhai_PRB2011,Baym_PRL2012}. 
Here, to study strongly interacting lattice bosons, we formulate a stochastic
Gutzwiller approach, which treats strong quantum correlations at mean field level, but retains full knowledge of spatial thermal fluctuations. 
The Monte Carlo (MC) technique introduced here is of broad applicability, being especially useful when the sign 
problem prevents quantum MC simulations, such as for frustrated bosons.
(iv) Using this approach, we obtain the concrete temperature-doping phase diagram of spinor lattice bosons with SOC 
and strong correlations as shown in Fig.~2. Thermal fluctuations are shown to destroy superfluidity well below the stripe transition, 
leading to a wide window of a normal Bose fluid with stripe order, this regime being enhanced near the Mott insulator.

\begin{figure*}[t]
\subfloat[]{\includegraphics[scale=0.20]{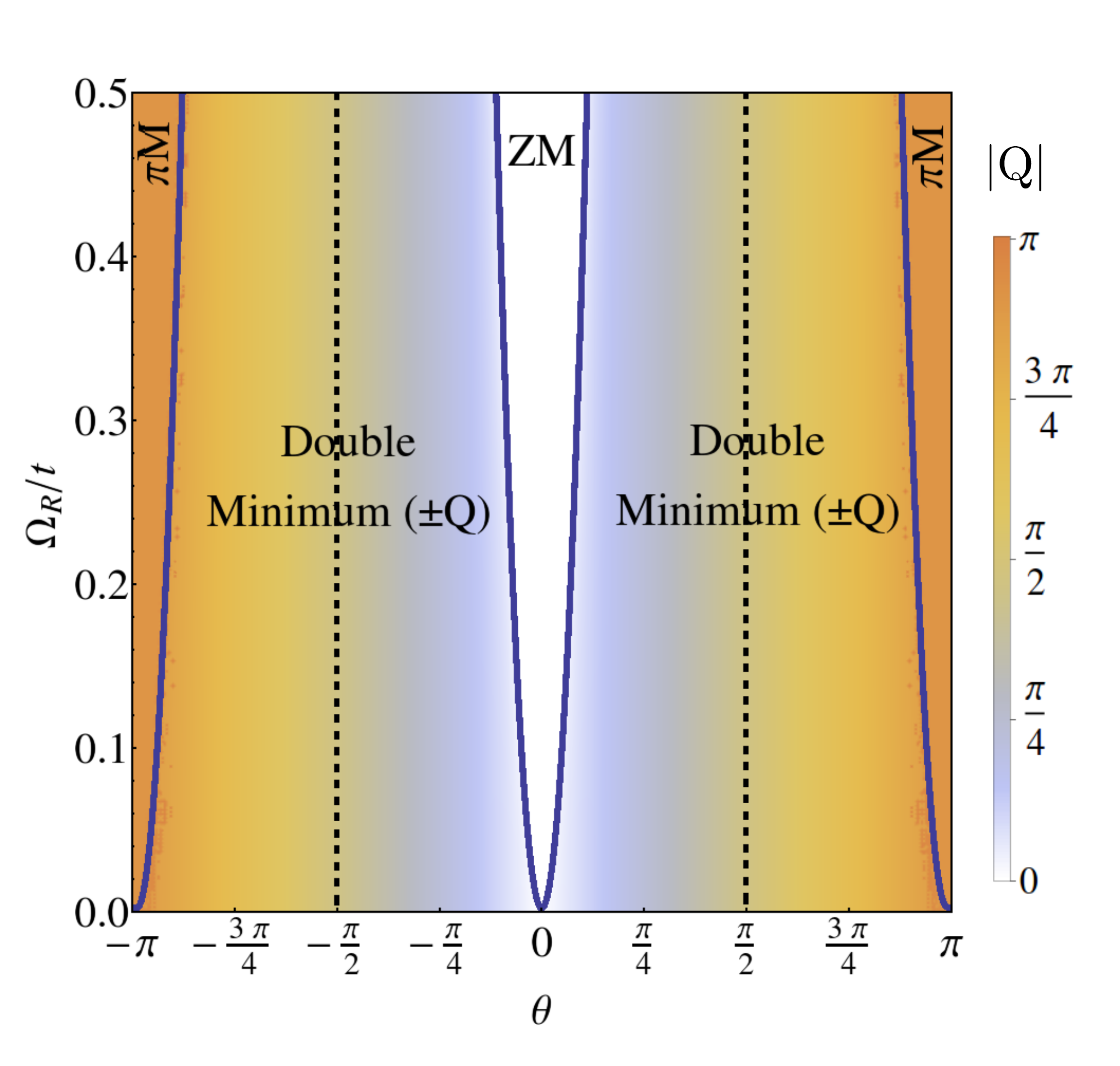}}
\subfloat[]{\includegraphics[scale=0.20]{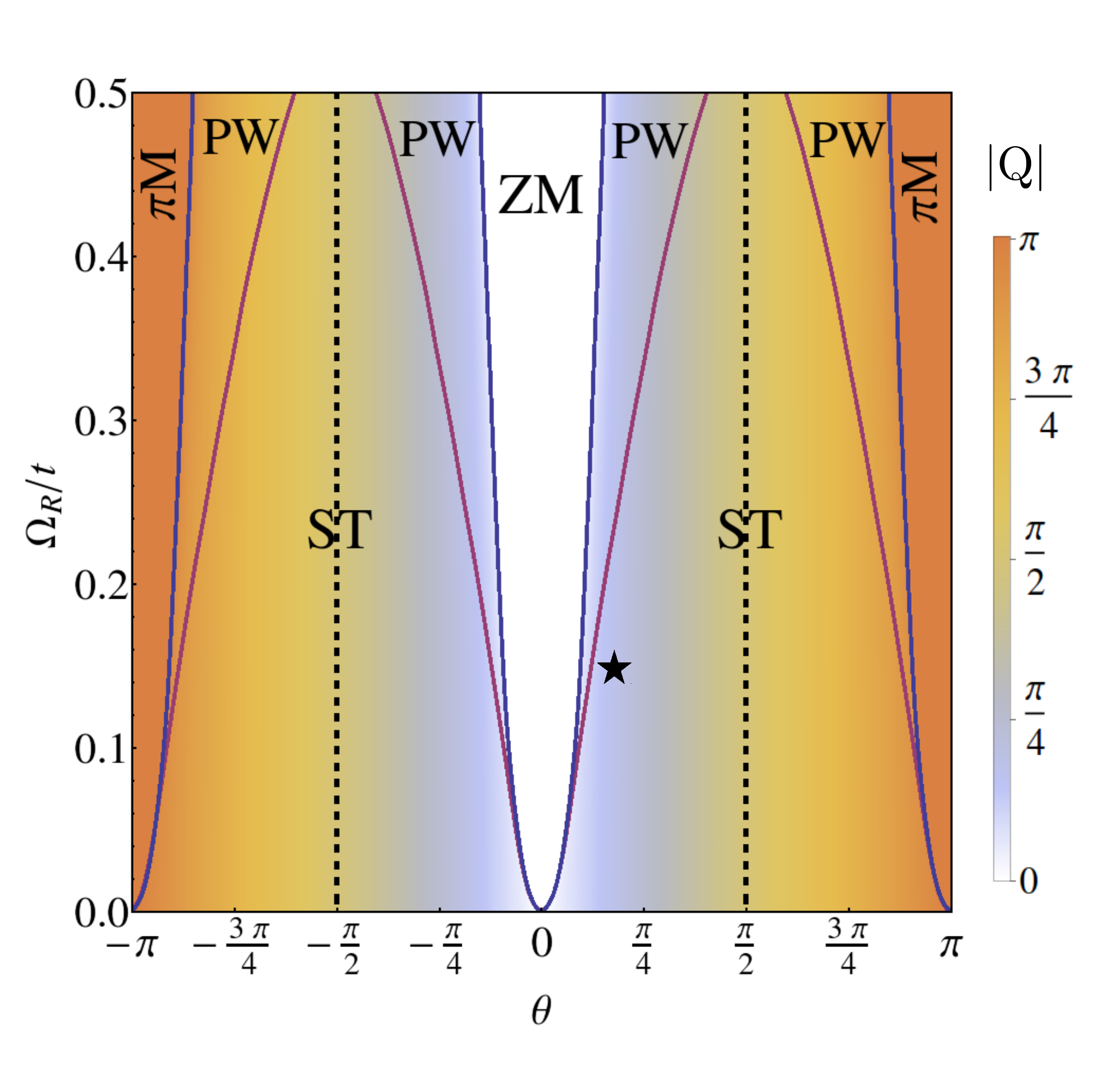}}
\subfloat[]{\includegraphics[scale=0.135]{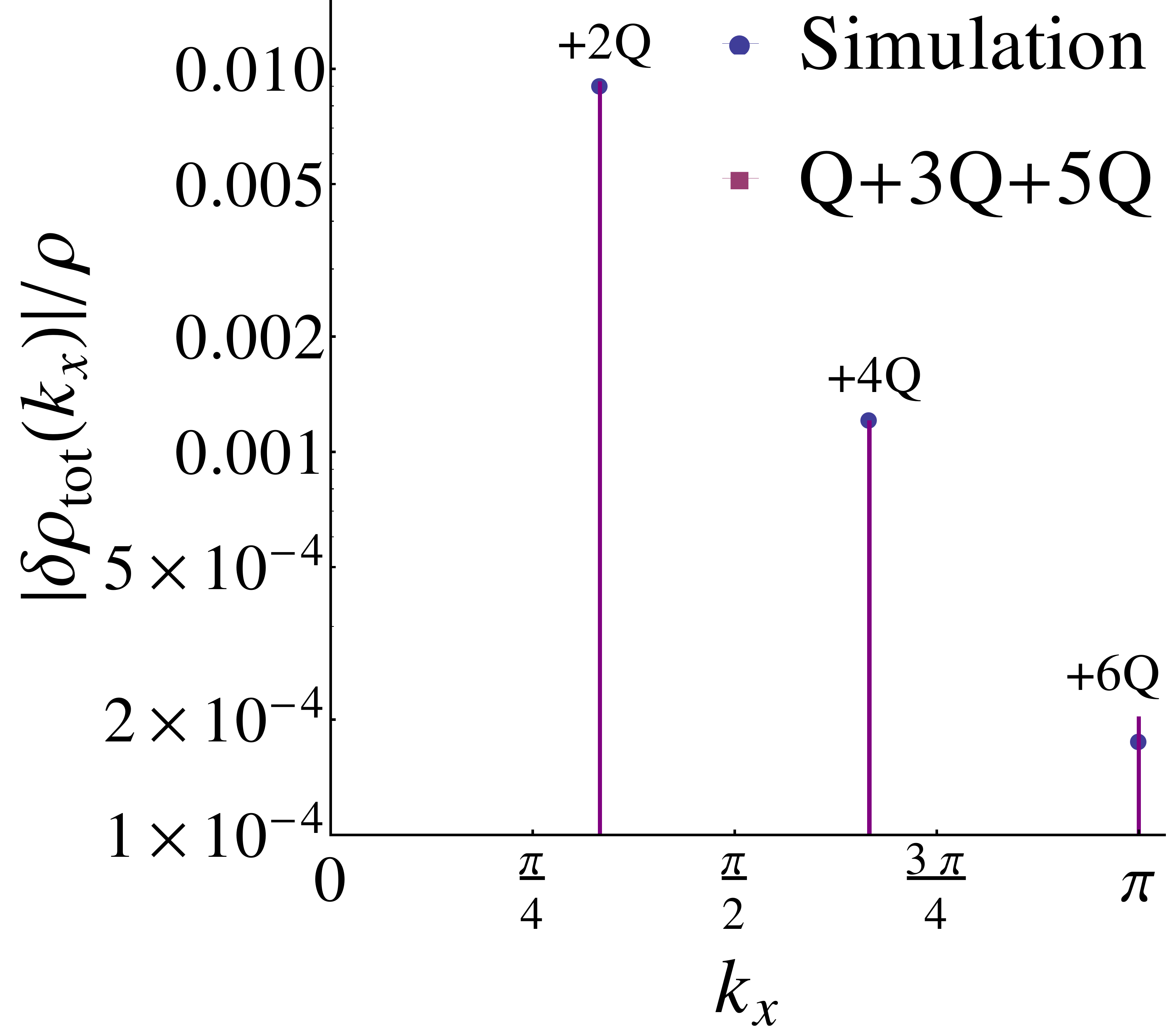}} \,\,
\subfloat[]{\includegraphics[scale=0.185]{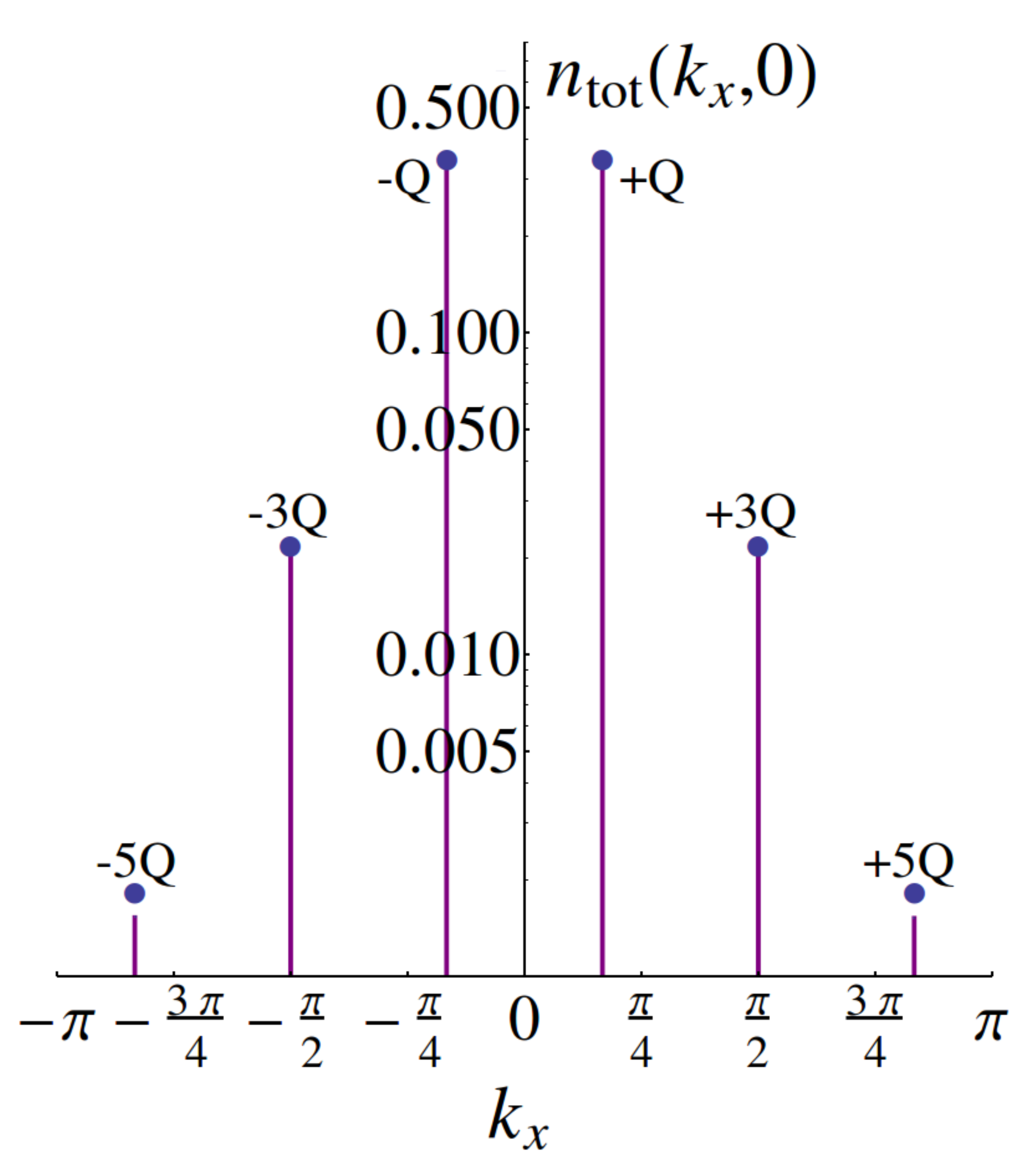}}
\caption{(a) Magnetic field ($\Omega_R$) evolution of the dispersion for
noninteracting bosons, with double minima at $(\pm Q,0)$ for SOC angle 
$\theta$. At large $\Omega_R$, we find a unique minimum at $Q=0$ (i.e., $ZM$) or $Q=\pi$ (i.e., $\pi$M). At $\theta=\pm\pi/2$, minima are pinned to $(\pm \pi/2,0)$. (b) 
Interacting $T=0$ phase diagram at density $\rho=0.5$, showing emergent
plane wave ($PW$) and stripe ($ST$) states at $U/t=10$ and $\lambda=0.95$. $\star$ is the point at which we plot
(c) density modulation, and (d) momentum distribution, comparing simulations (dots) and the variational ansatz (line)
with three harmonics.}
\label{fig:Phase_Diagrams}
\end{figure*}

{\it Noninteracting lattice Hamiltonian. ---}
We work on a square optical lattice with lattice spacing $d$, and consider the hopping Hamiltonian for two-component bosons,
\bea
\!\!\!\! H_{\rm kin} \!=\! -t \! \sum_{\la ij \ra} (b_{i \alpha}^{\dagger}R_{\alpha\beta}^{i,j}b^\pdg_{j\beta} + h.c.)
- \frac{\Omega_R}{2} \! \sum_i  \left( n_{i\upspn}  \!-\! n_{i\dwnspn} \right). \label{eqn:Lattice_H}
\eea
Here, $R^{i,i+\hat{y}} = \mathbb{I}$, $R^{i,i+\hat{x}} = e^{i\theta \sigma_y}$, and the SOC angle $\theta$ dictates the ratio of spin-flip to spin-conserving hopping
amplitudes. For long wavelength modes, with momenta $k \ll 1/d$, this Hamiltonian reduces to
\be
H^{\rm long}_{\rm kin}(\bk) \approx b^\dagger_{\bk\alpha} \left[\frac{k_\ell^2}{2 m_\ell} \delta^{\alpha\beta} + \gamma k_x \sigma^{\alpha\beta}_y -  
\frac{\Omega_R}{2} \sigma^{\alpha\beta}_z \right]
b^\pdg_{\bk\beta},
\ee
where $\ell=x,y$ with an implicit sum over $\ell$.
This is the form of the experimentally realized continuum Hamiltonian (at zero detuning). We identify the equal Rashba-Dresselhaus SOC coupling $\gamma=2 t d \sin\theta$,
anisotropic inverse effective masses, $m^{-1}_x \!=\! 2 t d^2 \cos\theta$ and $m^{-1}_y \!=\! 2 t d^2$, induced by the lattice, and a Raman laser induced Zeeman field
$\Omega_R$. (Henceforth, we set $d=1$.)
For general $\bk$, we find mode energies on the lattice
\be
\!\! E^\pm_\bk \! =\! - 2 t \left( \cos\!\theta\! \cos \! k_x \!+\! \cos \! k_y \right) \! \pm\! \sqrt{\frac{\Omega_R^2 }{4} \!+\! 4 t^2\! \sin^2\!\theta \sin^2\!k_x}. \nonumber
\ee
Focusing on the lower branch, $E^{-}_\bk$, the dispersion exhibits degenerate minima at $(k_x,k_y)=(\pm Q,0)$, similar to the continuum. 
At $\Omega_R\!=\!0$, 
we get $Q\!=\! \theta$. 

For $\Omega_R \neq 0$, we find three regimes. (i) $-\pi/2 \!<\! \theta \!<\! \pi/2$: Increasing $\Omega_R$ leads to $Q \!\to\! 0$, and we eventually lock into $Q=0$
for $\Omega_R > \Omega^c_R \equiv 4 t |\sin\theta\tan\theta|$; this regime is labelled zero-momentum (ZM). (ii)
$\pi/2 \!<\! |\theta| \!<\! \pi$: The minima shift in the opposite direction with increasing field, locking into $Q\!=\!\pi$ for $\Omega_R > \Omega^c_R$, a regime
we label $\pi$-momentum ($\pi$M).  A similar $\pi$M state, but with $k_y=\pm \pi$, is found for bosons with Rashba SOC in a 1D
{\it spin-dependent} periodic potential along the $x$-direction \cite{Han2012}, which acts as a staggered magnetic field.
(iii) $\theta\!=\! \pm\pi/2$: Here an extra
symmetry appears, namely,
${\cal U}^\dagger H_{\rm kin} {\cal U}\!=\! H_{\rm kin}$,
where the unitary operator ${\cal U}$ acts as ${\cal U}^\dg b^\pdg_{i\alpha} {\cal U}\!=\! b^\pdg_{{\cal M}_i \alpha} (-1)^{x_i}$, with
the site ${\cal M}_i \!\equiv\! (-x_i,y_i)$ obtained by reflection about the $y$-axis.
In momentum space, this sends $k_x \!\to\! -k_x \!+\! \pi$, which maps the minimum back onto itself, pinning $Q$ to
$\pm \pi/2$ for any $\Omega_R$. The
strong field limit on the lattice
thus leads to richer possibilities than the continuum \cite{Galitski2008,Zhai2010,Ho2011,Radic_PRA2011,LiPitaevskii_PRL2012,MartoneLiPitaevskii_PRA2012,Zhai2012,LiPitaevskii_PRL2013,Sheehy2013}.
Fig.~1(a) depicts the dispersion as a function of $\Omega_R/t$ and
$\theta$, tracking the evolution of $Q$, and marking boundaries where we reach $Q=0,\pi$.
A degenerate ``double well'' in the dispersion at $(\pm Q,0)$ leads to a macroscopic degeneracy of many-body ground states for
noninteracting bosons.
We next study how strong correlation effects break this degeneracy.

{\it Strongly interacting regime. ---} The hopping Hamiltonian in Eq.~\ref{eqn:Lattice_H} is in the conventional gauge choice where the atomic hyperfine states are
eigenstates of $\sigma_y$. Labelling hyperfine flavors by $c,d$, the local Hubbard interaction $H_{\rm U}=U_{cc} n_c (n_c-1)/2 + U_{dd} n_d (n_d-1)/2
+U_{cd} n_c n_d$. We choose $U_{cc}=U_{dd}=U$ and set $U_{cd}=\lambda U$ ($\lambda < 1$ for $^{87}$Rb). For $U \gg t$, double occupancy of bosons 
leads to a large energy 
cost. For fillings $\rho \leq 1$ boson per site, we thus use perturbation theory in $t/U$ \cite{WSCole_PRL2012, RadicGalitski_PRL2012, ZiCai_PRA2012} to
derive an effective Hamiltonian in the restricted Hilbert space where double occupancies are forbidden 
(see Supplemental Material \cite{SuppMat} for derivation).
The resulting effective strong coupling Hamiltonian is given by
\be
H_{\rm eff} = {\cal P} H_{\rm kin} {\cal P} \!+\! \sum_{i\delta} J_{\delta}^a S_i^a S_{i+\delta}^a + \sum_{i\delta} \vec{D}^\pdg_{\delta} \cdot (\vec{S}^\pdg_i \times 
\vec{S}^\pdg_{i+\delta})
\label{eqn:PertH}
\ee
with $\delta=\hat{x},\hat{y}$.
The first term denotes the kinetic energy term in 
Eq.~\ref{eqn:Lattice_H} (including the magnetic field $\Omega_R$) projected to the Hilbert space of no double occupancy, with ${\cal P}$ being the 
Gutzwiller projection operator. The next two terms describe
exchange interactions, with the spin operator $\vec S^a_i = \frac{1}{2} b^\dg_{i\alpha} \sigma^a_{\alpha\beta} b^\pdg_{i\beta}$, and
the exchange coefficients $J^a_{\delta}$ and Dzyaloshinskii-Moriya vectors $\vec D_\delta$ listed Table \ref{table:J_Terms}.

{\it Zero temperature phase diagram. ---}
The Gutzwiller ansatz provides a powerful approach to strongly correlated bosons \cite{Gutzwiller_PRL1963,Krauth1992}.
 This variational wavefunction is constructed as a
direct product (over all sites) of single-site wavefunctions, with each single-site wavefunction being capable of describing states with fluctuating 
or fixed particle number, thus providing a mean field description of a superfluid or a Mott insulator ground state. For two-component bosons \cite{WSCole_PRL2012}
the ansatz including the spin degree of freedom and no double occupancy constraint is
\begin{equation}
 \ket{\Psi} =  \overset{N}{\underset{i=1}{\otimes}} ( \chi_{i0} \ket{0} + \chi_{i\upspn}  \ket{\upspn} + \chi_{i\dwnspn} \ket{\dwnspn})  \label{eqn:GWF}
\end{equation}
where $\chi_{i n}$ are complex variational parameters, with normalization fixing $\sum_n \abs{\chi_{i,n} }^2 =1$ at each site $i$ (with $n=0,\upa,\dna$).
Minimizing $\la \Psi| H_{\rm eff} | \Psi \ra$ by optimizing $\{\chi_{i n}\}$ yields the phase
diagram shown in Fig.~1(b).

We highlight three key differences between the lattice phase diagram and its continuum counterpart. (i) The single-particle dispersion has two degenerate 
minima at $\bk=(\pm Q,0)$; this leads to a macroscopic many-body ground state degeneracy for noninteracting bosons, since they
can condense into any arbitrary superposition of wavefunctions 
constructed from these minima. Interactions split this degeneracy resulting in two phases for $\lambda<1$: a Stripe ($ST$) state featuring an equal superposition of the 
two minima, and a Plane Wave ($PW$) featuring condensation
into a single minimum. However, the lattice features two distinct $ST$ and $PW$ phases, with momentum distribution peaks
evolving with $\Omega_R$ to be closer to $ZM$ or $\pi M$. In addition, the wavevector of the $ST$ state at $\theta \! =\! \pm\pi/2$ is pinned to $Q=\pm \pi/2$ 
at all $\Omega_R$, since interactions preserve the previously discussed symmetry, so that
${\cal U}^\dagger H_{\rm eff} {\cal U}= H_{\rm eff}$. (ii) Strong correlations suppress $\Omega_R^c$
by a factor $\!\sim \!(1\!-\! \rho)$, leading to an enlarged window of $ZM$/$\pi M$ (see Fig.~1(a,b)).
(iii) The continuum $ST$ state has a density
modulation with a dominant harmonic amplitude $\delta\rho(2Q) \! \sim \! m_z$, where $m_z$ is the uniform magnetization induced by $\Omega_R$
\cite{LiPitaevskii_PRL2012}.
By contrast, the lattice $ST$ state has strong mode-mode coupling,
lead to higher order Fourier peaks in the density and momentum distribution; see Fig.~1(c,d). This
suppresses the real space density modulation by an order of magnitude, while still allowing for significant $m_z$.

\begin{table}
\caption{\label{table:J_Terms} Exchange couplings along the $\hat{x},\hat{y}$ directions in the strong coupling bosonic $tJ$ Hamiltonian in Eq.\ref{eqn:PertH}}
\begin{ruledtabular}
\begin{tabular}{l l}
$J^x_{\hat{x}} =  -\frac{4t^2}{\lambda U} \cos2\theta$ &  $J^x_{\hat{y}}  = -\frac{4t^2 }{\lambda U}$ \\
$J^y_{\hat{x}}  = -\frac{4t^2}{\lambda U} ( 2 \lambda - 1 )$ & $J^y_{\hat{y}}  = -\frac{4t^2 }{\lambda U} ( 2 \lambda - 1 )$ \\
 $J^z_{\hat{x}}  =-\frac{4t^2}{\lambda U} \cos2\theta$  & $J^z_{\hat{y}}  =  -\frac{4t^2}{\lambda U}$ \\
$\vec{D}_{\hat{x}} =  -\frac{4t^2}{\lambda U} \sin2\theta \,\hat{y}$ & $\vec{D}_{\hat{y}} = 0$ \\
\end{tabular}
\end{ruledtabular}
\end{table}

The various phases we find from our numerical minimization are reasonably captured by  a variational ansatz
\bea
\!\!\! \begin{pmatrix} \chi_{i \upspn} \\ \chi_{i \dwnspn} \end{pmatrix} \!\! =\!\!\!\!\! \sum_{n={\rm odd}}  \!\!
\left[ \! \frac{c_{n}}{\sqrt{2}} \begin{pmatrix} \! a_{n} \! \\ \! b_{n} \! \end{pmatrix} \! e^{i Q n x_i} 
\! +\!   \frac{c_{-n}}{\sqrt{2}} \begin{pmatrix} \! a_{n} \! \\ \! - b_{n} \! \end{pmatrix} \! e^{-i Q  n x_i} \! \right]
\eea
where $a_n \!=\! \sin\phi_n \!+\! \cos\phi_n$, $b_{n}\!=\! i (\sin\phi_n \!-\! \cos\phi_n)$, the sum is over
odd integers $n\!>\! 0$, and $\chi_{i 0} \!=\! (1 \!-\! \abs{\chi_{i\upspn}}^2 \!-\! \abs{\chi_{i\dwnspn}}^2)^{1/2}$.  
Retaining the leading term ($n\!=\! 1$) reveals three states: $(i)$ Stripe ($ST$) order 
with $c_1=c^*_{-1}=\sqrt{\rho/2}$, representing an equal superposition of modes at $(\pm Q,0)$, $(ii)$ Plane Wave ($PW$)
order with $\{c_1, c_{-1}\}=\{0,\sqrt{\rho}\}$ or  $\{\sqrt{\rho},0\}$ representing a single mode condensate at
$(\pm Q,0)$, 
and $(iii)$ a $ZM/\pi M$ state with spins  fully polarized along the $\Omega_R$-axis. Limiting to $n\!=\!1,3,5$ quantitatively 
captures the leading harmonics in the density and momentum distribution in Fig.~1(c,d), but leads to
a 20\% error in the highest harmonic resolved in our simulations.

A strong coupling perspective is afforded by the local gauge transformation, 
$b_i = \left( b_{i\upspn}, b_{i\dwnspn} \right)^T \goto e^{ -i \theta x_i \sigma_y } \tilde{b}_i$, which leads to
\begin{align}
\nonumber \tilde{H}_{\rm eff}  = &  - t \sum_{\avg{i j}} \left( \tilde{b}_{i\alpha}^\dagger \tilde{b}_{j\alpha} + h.c. \right)  + \sum_{\la i j \ra}  \tilde{J}^\alpha
\tilde{S}^\alpha_i \tilde{S}^\alpha_j     \\
& - \Omega_R \sum_i   \left( \cos(2\theta x_i) \tilde{S}_i^z  -  \sin(2\theta x_i) \tilde{S}_i^x \right)
\end{align}
where $\tilde{J}^x=\tilde{J}^z=-4 t^2/\lambda U$ and $\tilde{J}^y= ( 1- 2 \lambda) 4 t^2/\lambda U$. We will assume $\lambda < 1$.
For $\Omega_R=0$ the spins align ferromagnetically in the $\hat{x}$-$\hat{z}$ plane. Such a state corresponds to $ST$ order in the original gauge. 
Large $\Omega_R$ forces spins to align with the local field; this is the $ZM$ state in the original gauge. At small $\theta$, aligning with this
external field does not cost much exchange energy since the spiral has a large pitch,
so the critical $\Omega^c_R$ is small. However, at larger $\theta$,
the exchange cost disfavors alignment with the spiralling field; instead, those spins parallel to the applied field simply increase their magnitude
by a local density enhancement at the expense of those antiparallel to the field, leading to a density modulated stripe.
At larger $\Omega_R$, spins flip out of the $\tilde{S}_x$-$\tilde{S}_z$ plane,  forming 
a `cone' state around the $\tilde{S}_y$ axis. This corresponds to the $PW$ state. The cone angle
grows with $\Omega_R$, eventually leading to a $ZM$ state. This sequence corresponds to a 
first order $ST$-$PW$ transition as $\tilde{S}_y$ suddenly becomes non-zero, followed by a continuous transition to $ZM$ order. 

\begin{figure*}[t]
\subfloat[]{\includegraphics[scale=0.13]{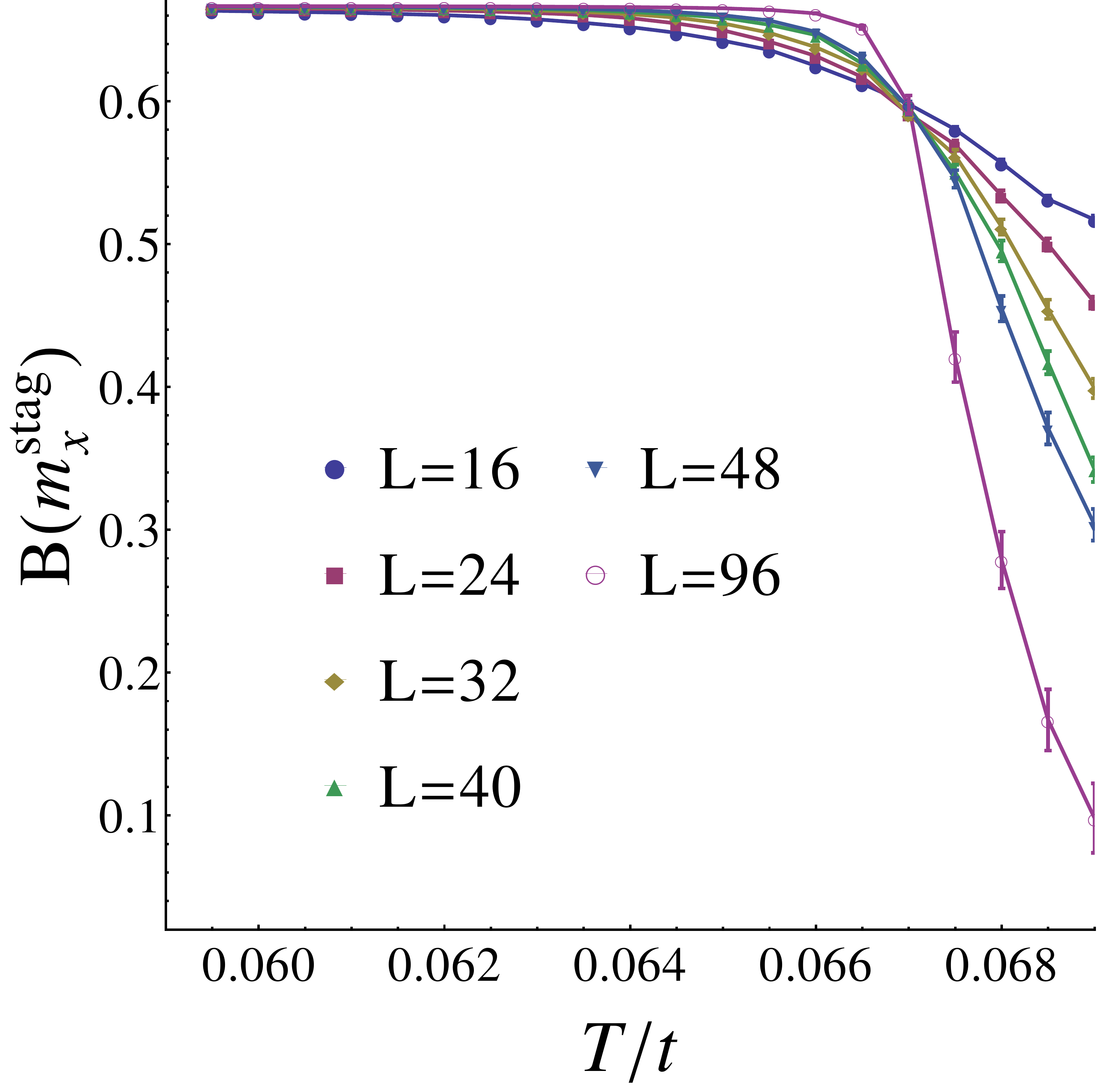} \label{fig:Binder}}
\subfloat[]{\includegraphics[scale=0.17]{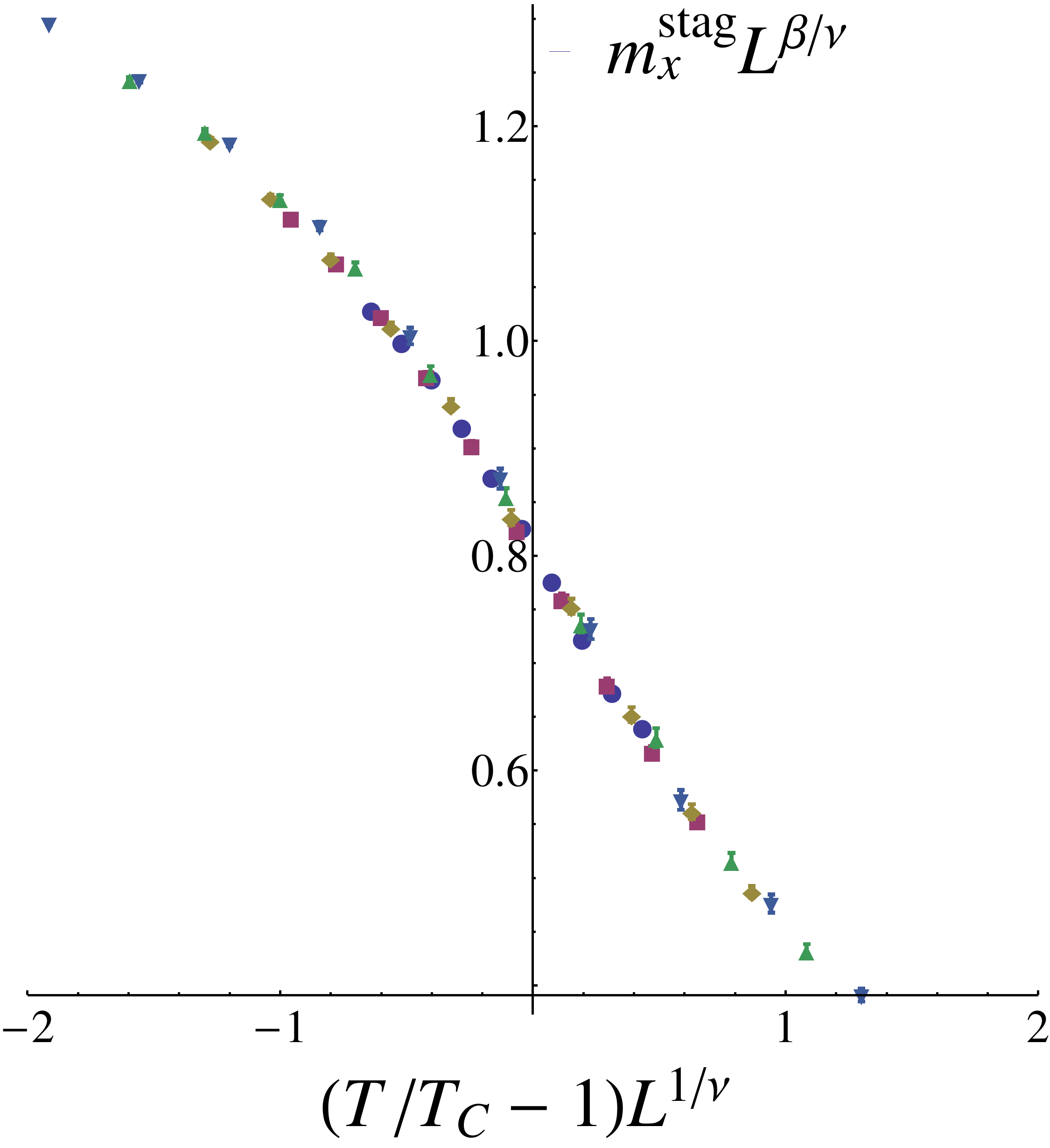}}
\subfloat[]{\includegraphics[scale=0.13]{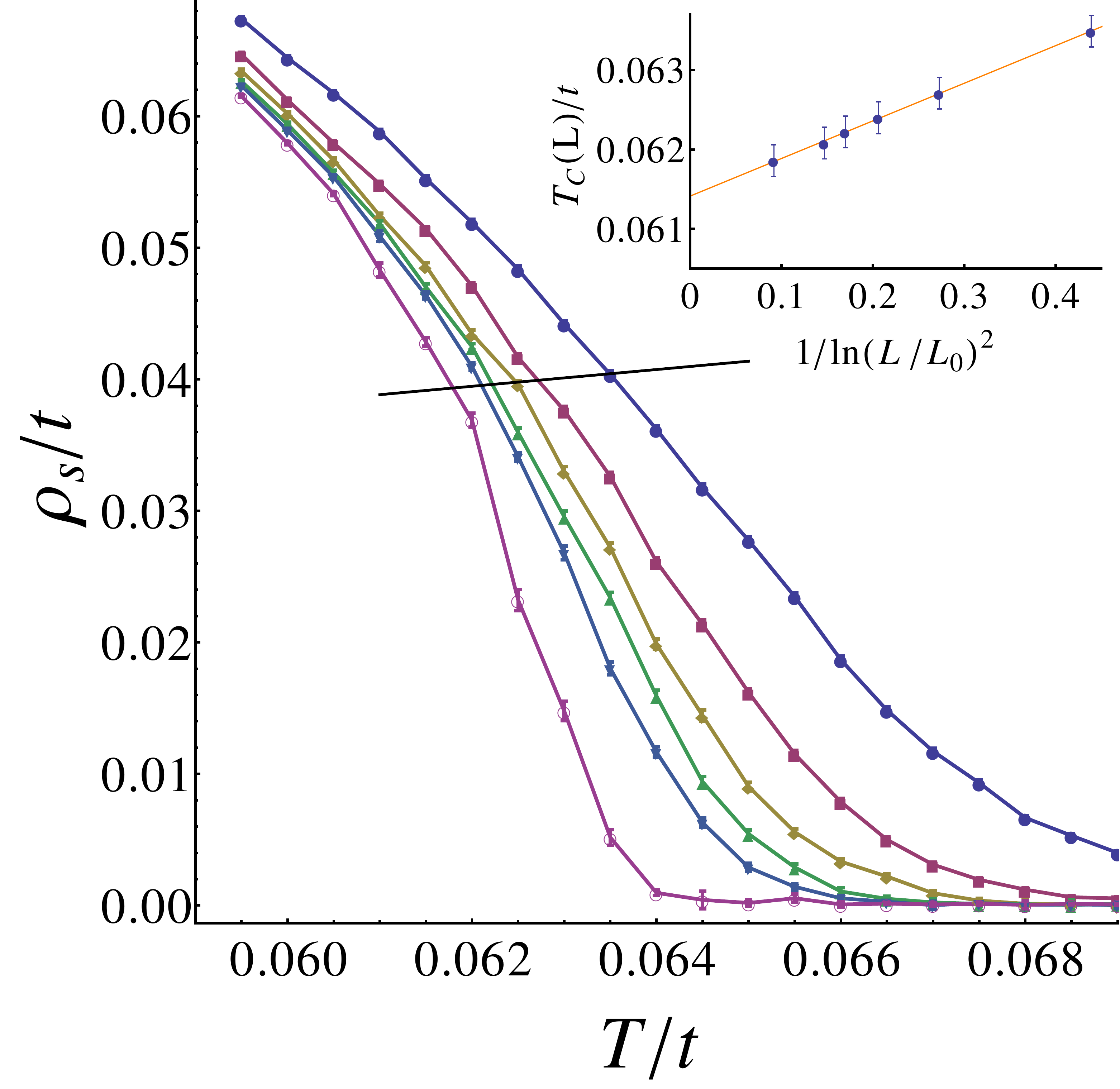} \label{fig:n_k}}
\subfloat[]{\includegraphics[scale=0.11]{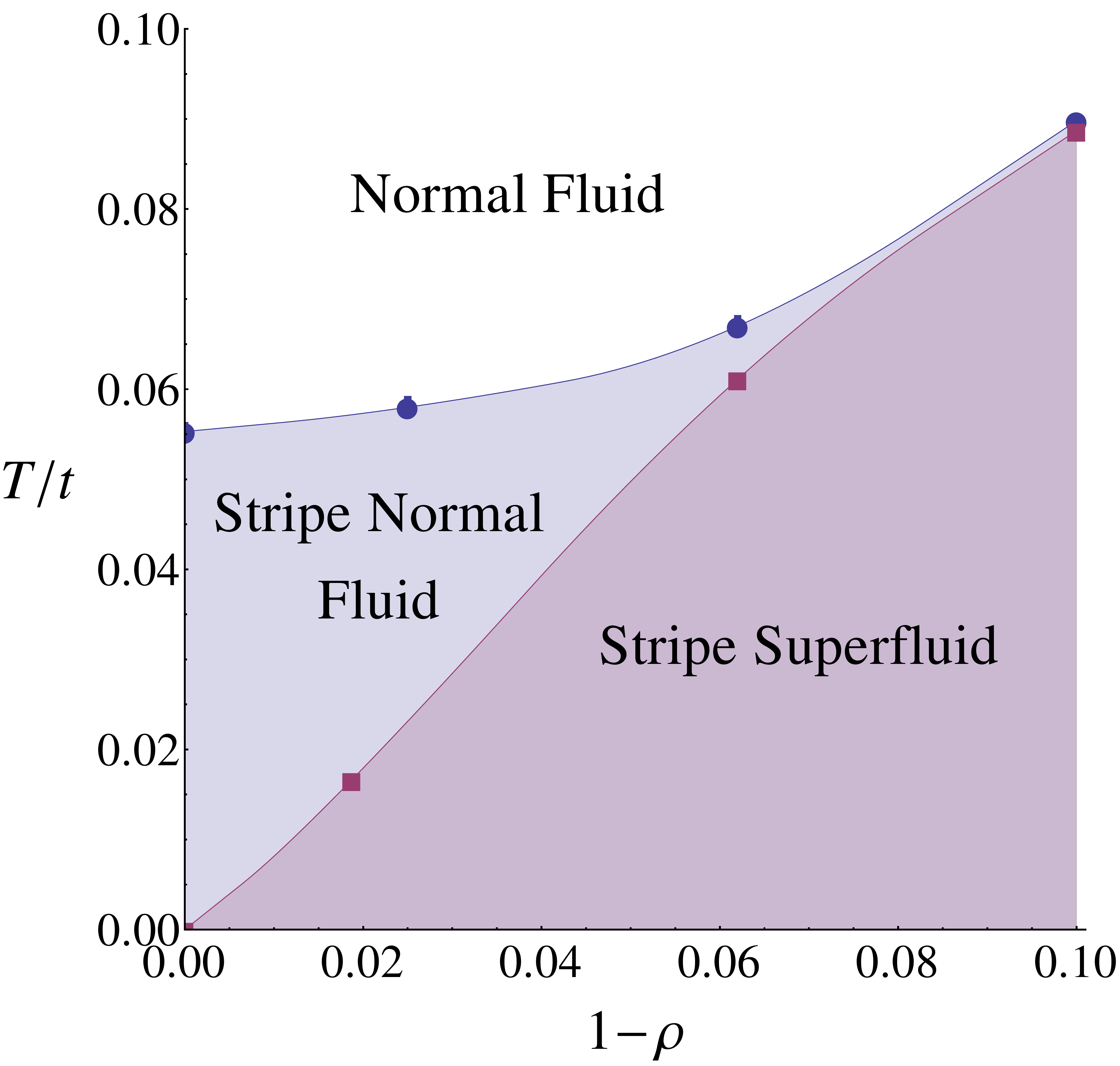}}
\caption{(a) Binder cumulants of the staggered magnetization $m^{\rm stag}_x$ for $L\times L$ systems at $\theta=\pi/2$, with
$U\!=\!10t $, $\rho\!=\!0.94$, and $\Omega_R\!=\! 0.5 t$,
showing crossing at $T_{\rm Ising}\!=\! 0.067(1) t$.
(b) Scaling collapse of the order parameter $m^{\rm stag}_x$ using Ising exponents $\beta\!=\! 1/8$ and $\nu\!=\! 1$. (c) Superfluid
stiffness $\rho_s(T)$, showing finite size transition temperature $T_c(L)$ given by crossing of $\rho_s(T)$ with the line $2 T/\pi$. Inset shows
extrapolation of $T_c(L)$ to the thermodynamic limit, yielding $T_{\rm BKT}\!=\! 0.0614(2) t$. (d) Doping-temperature phase diagram showing the
emergence of a stripe normal phase.}
\label{fig:Thermal}
\end{figure*}

{\it Thermal fluctuations and transitions. ---}
To study strong correlations at nonzero temperature $T=1/\beta$, we express the partition function 
$Z = \text{Tr} \left( e^{-\beta H} \right)$ in path integral form using the basis of Gutzwiller wavefunctions,
\be
Z \!=\!\! \int \! {\cal D}\chi\chi^* \bra{ \Psi} e^{ - \beta H_{\rm eff}} \ket{ \Psi} \!\! \approx \!\! \int \! {\cal D} \chi\chi^* e^{- \beta  \bra{ \Psi} H_{\rm eff} \ket{ \Psi}},
\ee
where the final approximation uses the leading order term in a cumulant expansion. This cumulant approximation  \cite{Stoudenmire2009} is 
exact at $T=0$, recovering the ground state energy with mean field quantum correlations, and is also exact to
leading order in $1/T$ in a high temperature expansion (see Supplemental Material \cite{SuppMat} for details).
We thus expect this approximation to accurately capture thermal fluctuation effects over the entire range of
temperatures.

To compute physical observables, we use a Monte Carlo approach to sample the partition function and calculate observables, treating $\chi_{i,n}$ as stochastically
fluctuating variables. 
This method generalizes in a straightforward manner
if we relax the no double-occupancy constraint to allow for a maximum occupancy $n_{\rm max}$ bosons 
at each site including both species. In this case, each site has a complex vector of $(n_{\rm max}+1)(n_{\rm max}+2)/2$ fluctuating 
components. Since there is no sign problem, this method is also suitable for studying thermal fluctuations in frustrated
bosons and their Mott transitions.

For generic $\theta$, the Bose condensation wavevector and the magnetic order will be incommensurate, and will shift with $\Omega_R$ and $T$. 
This makes it numerically more difficult to accurately locate the thermal transitions. Here, we therefore illustrate this
method by studying the effect of thermal fluctuations at $\theta=\pi/2$, which ensures that the ordering wavevector $Q= \pi/2$ is independent
of $\Omega_R$ and $T$, enabling us to precisely locate the thermal transitions.

At $\theta=\pi/2$, the staggered magnetization, $m_x^{stag} \equiv \sum_i \left( -1 \right)^{x_i} \la S^x_{i} \ra$, breaks $Z_2$ symmetry when $\Omega_R \neq 0$. To probe the transition where magnetism is lost, we compute the Binder cumulant \cite{Binder_ZPBCM1981} curves of the order parameter. As shown in Fig.~2(a), for 
$U/t=10$, $\rho=0.94$, $\Omega_R=0.5 t$,
these show a unique crossing
point, which allows us to locate $T_{\rm Ising} = 0.067(1) t$. We find, as shown in Fig.~2(b), that the scaled
order parameter near
$T_{\rm Ising}$ collapses onto a single curve for Ising exponents, $\beta=1/8$ and $\nu=1$.

We track the destruction of superfluid order by computing the superfluid stiffness. Since the Hamiltonian is anisotropic in space,
the stiffness is different along $\hat{x}$ and $\hat{y}$, and the geometric mean $\rho_s \!=\! \sqrt{\rho_s^{xx} \rho_s^{yy}}$ controls the energy of 
vortices which proliferate and destroy superfluidity. As seen in Fig.2(c), $\rho_s$ drops rapidly with temperature reminiscent of the behavior near
a Berezinskii-Kosterlitz-Thouless (BKT) transition. We confirm this by identifying the finite size superfluid transition temperature $T_c(L)$ via the 
intersection point defined by $\rho_s \left( T_c(L) \right) \!=\! 2 T_{c}(L) /\pi$, and finding that $T_c(L)$ obeys the expected scaling form 
$T_C (L) \!=\! T_{\rm BKT}  \!+\! b/\ln^2 (L/L_0)$ (see Fig.2(c) inset),  where $b$ and $L_0$ are non-universal numbers. This also allows us to extract the thermodynamic
limit transition temperature $T_{\rm BKT} \!=\! 0.0614 (2) t$. We have confirmed the BKT nature of the transition
from the critical scaling of $n(\bf{k})$ (see Supplemental Material \cite{SuppMat}).

Using the above methods to extract $T_{\rm Ising}$ and $T_{\rm BKT}$ at various densities $\rho$ enables us to construct the phase diagram in Fig.~2(d).
In the Mott insulator, at $\rho=1$, we find a single (Ising) transition associated with magnetic ordering. Upon doping, the stripe magnetic order survives,
but in addition superfluidity appears with a low transition temperature. This leads to a wide window of normal stripe order. With 
increasing doping away from the Mott insulator, the two transitions get closer to each other, and the normal stripe order shrinks.

{\it  Discussion. ---} For lattice bosons with SOC, we have uncovered strongly correlated superfluid ground states 
distinct from the continuum. At $T \! \neq\! 0$, we have used a stochastic Gutzwiller approach to
show that the $ST$ superfluid phase undergoes multiple transitions, revealing an intermediate stripe normal phase which increases in
width as one approaches the Mott insulator. Going beyond our specific calculations, we expect that even for $\theta \! \neq \! \pi/2$, magnetic order will 
persist in the Mott insulator, whereas the superfluid transiton temperature $T_{\rm BKT}$ will vanish as $\rho\! \to \! 1$; thus,
the stripe normal phase will persist even in this generic case. Furthermore, even if the repulsion is not strong enough to drive Mott
insulators, we expect the window of normal stripe fluid to be maximal near $ \rho \sim 1$, and the stripe normal phase should also persist
in higher dimensions.
Our phase diagram could
be explored using atomic bosons with SOC in optical lattices \cite{Hammer2014}. The stripe normal fluid would display broadened momentum peaks simultaneously
at  $\pm Q$, visible in time-of-flight experiments. The spin order in the normal stripe fluid could be probed using Bragg scattering experiments
\cite{Corcovilos2009}, similar to recent detection of N\'eel correlations in the atomic Fermi-Hubbard model \cite{Hart2014}.

{\it Note added. ---} During completion of this manuscript we became aware of complementary work \cite{Radic2014} which discusses
magnetic instabilities of normal (uncondensed) spin-1/2 bosons in the continuum.

{\it Acknowledgements. ---} We thank I. Kivlichan, P. Engels, V. Galitski, S.-B. Lee, S. Natu, H. Pu, H. Zhai, and N. Trivedi for useful discussions. We acknowledge 
funding from NSERC of Canada. AP thanks the Aspen Center for Physics (Grant No. NSF PHY-1066293) for hospitality during the completion 
of this manuscript.



\bibliographystyle{apsrev4-1}


%
  
  \vfill\eject

\widetext

\section{appendix}

\subsection{Derivation of $tJ$ model for two-component bosons with SOC}
With the hyperfine flavours labelled by $c$ and $d$ the local Hubbard interaction is
\begin{equation}
H_{\rm U} = \frac{U_{cc}}{2} \sum_i n_{i,c} ( n_{i,c} - 1) + \frac{U_{dd}}{2} \sum_i n_{i,d} ( n_{i,d} - 1) + U_{cd} \sum_i n_{i,c}n_{i,d} , \label{eqn:H_0}
\end{equation}
Setting $U = U_{cc} = U_{dd} $, $\lambda U = U_{cd}$ and $U \gg t$, we restrict ourselves to a Hilbert space in which double occupancy of sites is forbidden. Using second order perturbation theory in $t/U$ we can derive an effective Hamiltonian for this restricted space with $H_{\rm U}$ given by Eq.~\ref{eqn:H_0} and the perturbation $H_{\rm kin}$ given by Eq.~\ref{eqn:Lattice_H}. Written in terms of hyperfine basis states
the perturbation 
$H_{\rm kin}$ is
\begin{align}
H_{\rm kin} =& -t \sum_i  \left( e^{i\theta} c_{i}^{\dagger}c_{i+\hat{x}} + e^{-i\theta} d_{i}^{\dagger}d_{i+\hat{x}}  + h.c.  \right) -t \sum_i  \left(  c_{i }^{\dagger} c_{i+\hat{y}} +  d_{i}^{\dagger} d_{i+\hat{y}}  + h.c.      \right)   - \frac{\Omega_R}{2}  \sum_i \left( c_{i}^{\dagger}d_{i} + d_{i }^{\dagger}c_{i} \right)   .   \label{eqn:latticeHHYP}
\end{align}
Using a two-site basis of degenerate states $\left[ \ket{c,c}, \ket{c,d}, \ket{d,c}, \ket{d,d} \right]$ the matrix form of the effective Hamiltonian for the $\hat{x}$-direction is
\begin{equation}
H^J_x = \begin{pmatrix}  -\frac{4t^2}{U}  &  0  &   0  &  0 \\
0  &  -\frac{2t^2}{\lambda U}  &  -\frac{2t^2}{\lambda U} e^{2 i \theta} & 0  \\
0 &  -\frac{2t^2}{\lambda U} e^{-2 i \theta}  &  -\frac{2t^2}{\lambda U}  & 0  \\
0 & 0 & 0 & -\frac{4t^2}{U}  \end{pmatrix} , 
\end{equation}
while in the $\hat{y}$ direction
\begin{equation}
H^J_{y} = \begin{pmatrix}  -\frac{4t^2}{U}  &  0  &   0  &  0 \\
0  &  -\frac{2t^2}{\lambda U}  &  -\frac{2t^2}{\lambda U} & 0  \\
0 &  -\frac{2t^2}{\lambda U}  &  -\frac{2t^2}{\lambda U}  & 0  \\
0 & 0 & 0 & -\frac{4t^2}{U}  \end{pmatrix} .
\end{equation}
These can be rewritten in terms of spin operators as:
\begin{align}
H^J = & \sum_i \sum_{ \delta =  \hat{x},\hat{y}}  \left( \sum_{a = x, y, z} J_\delta^a S_i^a S_{i+\delta}^a \right) + \sum_i \vec{D} \cdot (\vec{S}_i \times \vec{S}_{i+\hat{x}}),
\end{align}
where $S_i^x =  (  b_{i \upspn}^\dagger b_{i \dwnspn} +  b_{i \dwnspn}^\dagger b_{i \upspn}   ) / 2 , \, S_i^y =  -i(  b_{i \upspn}^\dagger b_{i \dwnspn} -  b_{i \dwnspn}^\dagger b_{i \upspn} ) / 2 $ and $S_i^z = ( n_{i \upspn} - n_{i \dwnspn} ) / 2$ and the exchange coefficients $J_\delta^a$ and Dzyaloshinskii-Moriya vectors are given in Table \ref{table:J_Terms}. The total Hamiltonian is then given by ${\cal P} H_{\rm kin} {\cal P} + H^J$, as given in Eq.~3 of the paper.

\subsection{Details of finite temperature Gutzwiller method.}
Using the basis of Gutzwiller wavefunctions the partition function can be written as 
\begin{align}
\notag Z &= \text{Tr} \left( e^{-\beta H} \right) = \int {\cal D} \chi\chi^* \bra{ \Psi} e^{ - \beta H} \ket{ \Psi}, \\
& \approx \int {\cal D} \chi\chi^* e^{-\beta \bra{ \Psi}  H \ket{ \Psi} },
\end{align}
where in the last line we have approximated it by the leading order term in a cumulant expansion of the full partition function and the integration measure is 
\begin{equation}
{\cal D} \chi \chi^* = \prod_i \left[ \prod_n d\chi_{i,n} d\chi^*_{i,n} \right] \delta \left( \sum_n \abs{ \chi_{i,n} }^2 - 1 \right).
\end{equation}
where $0 \leq n \leq n_{\rm max}$.
Such a cumulant expansion has been used to study the appearance of quadrupolar correlations in 
a class of quantum spin-$1$ models in the literature \cite{Stoudenmire2009}.
At $T=0$ the approximation is exact, recovering the zero temperature Gutzwiller mean field result,
\begin{align}
\notag Z = \int {\cal D} \chi\chi^* e^{-\beta \bra{ \Psi_0}  H \ket{ \Psi_0} } =  \int {\cal D} \chi\chi^* e^{-\beta E_0 } =   \int {\cal D} \chi\chi^* \bra{ \Psi_0} e^{ - \beta H} \ket{ \Psi_0} .
\end{align}
Furthermore, at high temperatures we can expand the exponential 
\begin{align}
\notag Z &  \approx \int {\cal D} \chi\chi^* e^{-\beta \bra{ \Psi}  H \ket{ \Psi} } \approx \int {\cal D}\chi\chi^* \left( 1 - \beta \bra{ \Psi}  H \ket{ \Psi} +  \ldots\right),
\end{align}
which matches exactly the high temperature expansion of the full partition function to leading order in $1/T$
\begin{align}
\notag Z &= \int {\cal D} \chi\chi^* \bra{ \Psi} e^{ - \beta H} \ket{ \Psi}  \approx \int {\cal D} \chi\chi^* \bra{\Psi} \left( 1 - \beta H +  \ldots \right) \ket{\Psi}, \\
\notag  &  \approx \int {\cal D} \chi\chi^* \left( 1 - \beta \bra{ \Psi}  H \ket{ \Psi} +  \ldots\right) . 
\end{align}
We therefore expect this cumulant approximation to yield a good approximation to the full partition function and thermodynamic observables 
at all intermediate temperatures.

To sample the partition function, it is simplest to work in the grand canonical ensemble and
make local updates on $\chi_{i,n}$ by choosing any two components at a randomly chosen 
site and performing a random $SU(2)$ rotation on them which explicitly preserves the normalization. We choose the chemical potential to leave
the density fixed as we vary the temperature and magnetic field.

\begin{figure*}[tb]
\subfloat[]{\includegraphics[scale=0.18]{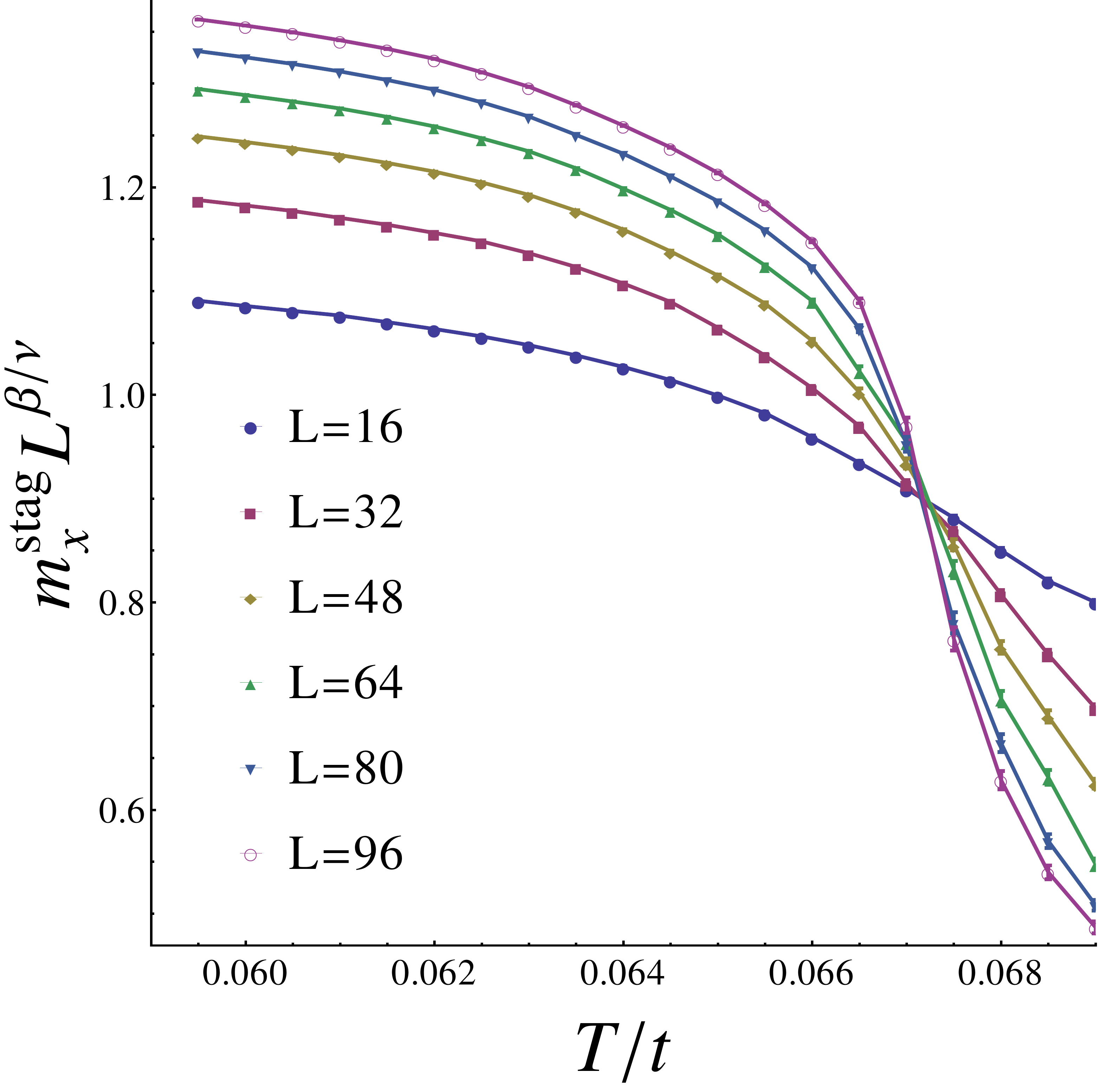}\label{fig:M_x_Plot} }\,\,\,\,
\subfloat[]{\includegraphics[scale=0.18]{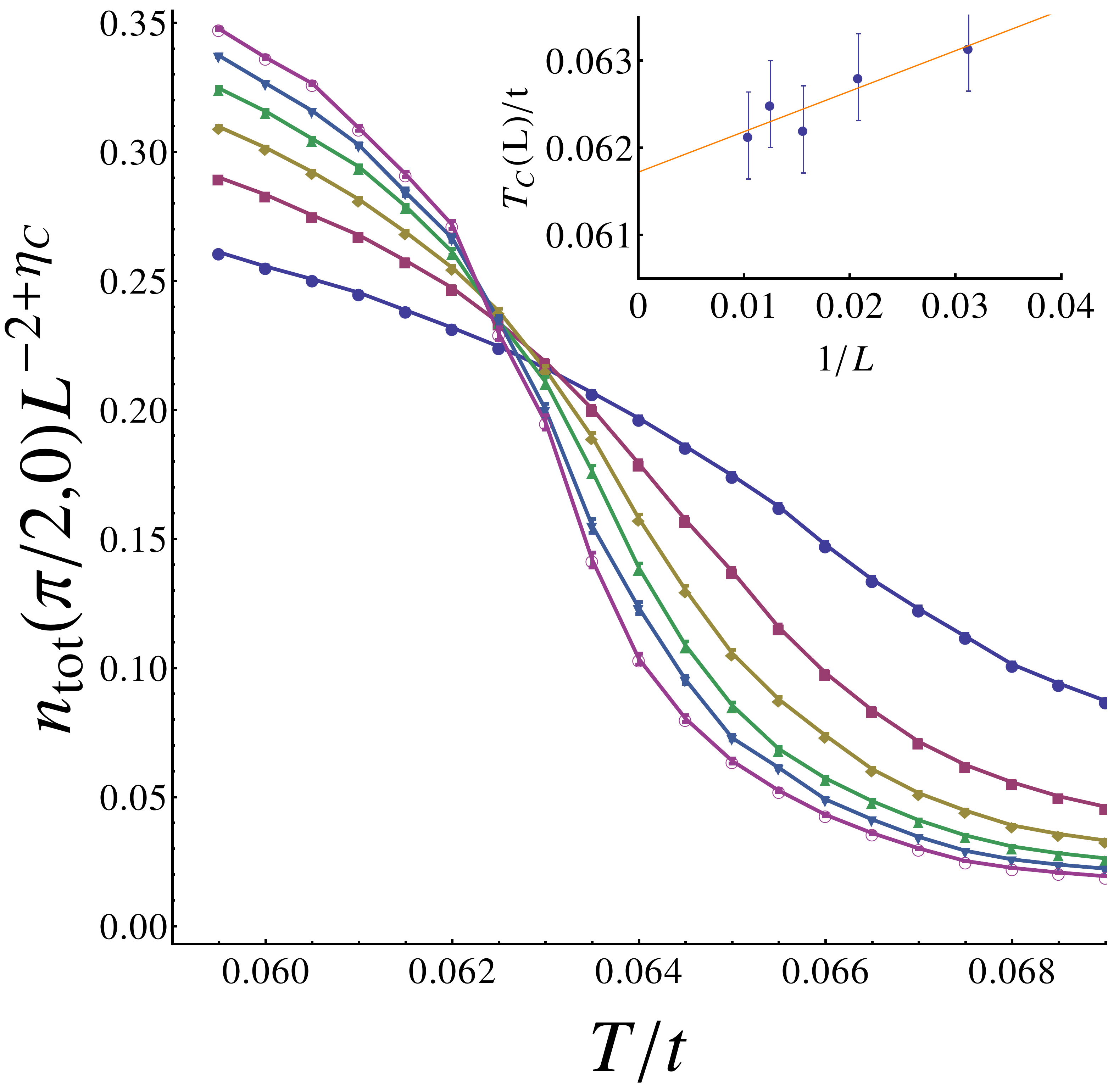} \label{fig:Mom_Dist_Plot}}
\label{fig:L74_Plots}
\caption{Plots of (a) the scaled 
staggered magnetization $m_x^{stag} L^{\beta/\nu}$ as a function of temperature $T$ for various system sizes $L\times L$,
and (b) the scaled momentum distribution $n_{\rm tot}(\pi/2,0) L^{-2 + \eta_C}$ (with $\eta_C=1/4$) for parameter values $U=10t, \lambda = 0.95, \Omega_R = 0.5t$ and a uniform density $\rho = 0.94$. }
\end{figure*}

\subsection{Confirmation of the nature of the thermal transitions}

{\bf Magnetic transition:}
We can obtain the magnetic transition temperature differently, by using the Ising nature of the magnetic critical point. 
We plot the scaled order parameter $m_x^{\rm stag} L^{\beta/\nu}$ with $\beta=1/8$ and $\nu=1$. There are three distinct behaviours expected for such a plot 
\begin{align}
\notag \text{ Disordered ($T > T_{\rm Ising}$):} \,\,\, m_x^{\rm stag} L^{\beta/\nu} &\sim L^{-1} L^{1/8} = L^{-7/8}, \\
\notag \text{Critical ($T = T_{\rm Ising}$):}\,\,\,  m_x^{\rm stag} L^{\beta/\nu} &\sim  L^{0}, \\
\text{Ordered ($T < T_{\rm Ising}$):}\,\,\,  m_x^{\rm stag} L^{\beta/\nu} &\sim L^{0} L^{1/8} = L^{1/8},
\end{align}
The curves are thus expected to cross at $T_{\rm Ising}$. The results are shown in Fig.~3(a) for $U=10t, \lambda = 0.95, \Omega_R = 0.5t$ and a uniform density $\rho = 0.94$, yielding $T_{\rm Ising} = 0.067 (1) t$, in agreement with the Binder cumulant result. 

{\bf Superfluid transition:}
To confirm the BKT nature of the superfluid transition we plot the scaled momentum distribution $n(\bk)L^{-2+\eta_C}$ at $\bk = \left( \pi/2, 0\right)$ for different systems sizes $L$, where $\eta_C = 1/4$ for a BKT transition. There are similarly three distinct behaviours expected
\begin{align}
\notag \text{ Disordered ($T > T_{\rm BKT}$):} \,\,\, n( \bk ) L^{ -2+\eta_C} &\sim L^0 L^{-7/4} = L^{-7/4}, \\
\notag \text{Critical ($T=T_{\rm BKT}$):}\,\,\,  n( \bk ) L^{ -2+\eta_C} &\sim  L^{0}, \\
\text{Algebraic Order ($T < T_{\rm BKT}$):}\,\,\,  n( \bk ) L^{ -2+\eta_C} &\sim L^{2 - \eta(T)} L^{-7/4} = L^{1/4 - \eta(T)},
\end{align}
The numerical results are shown in Fig.~3(b) for $U=10t, \lambda = 0.95, \Omega_R = 0.5t$ and a uniform density $\rho = 0.94$, with the crossing point clearly weakly drifting with system size $L$ due to logarithmic corrections to the
superfluid stiffness at the BKT transition.
In the inset, we plot the value of the crossing point for successive system sizes (called $T_c(L)$) as a function of $1/L$, where $L$ is the 
larger system size, which upon extrapolation to $L \to \infty$ yields $T_{\rm BKT} = 0.0617(2) t$, in agreement with the result obtained
from the superfluid stiffness calculation.

\end{document}